\documentclass[runningheads]{llncs}
\usepackage{graphicx}
\usepackage{amssymb, amsmath, bm, latexsym,comment}
\usepackage{multirow}
\usepackage{xcolor}
\usepackage{subfigure}
\usepackage{indentfirst}
\usepackage{setspace}
\usepackage{verbatim}
\usepackage{array}
\usepackage{arydshln}
\usepackage[misc]{ifsym} 
\usepackage[colorlinks,linkcolor=blue]{hyperref}

\providecommand{\Leireftb}[1]{Table~\ref{#1}}
\providecommand{\Leireffig}[1]{Fig.~\ref{#1}}

\providecommand{\citep}[1]{\cite{#1}}

\begin{document}
\title{AtrialGeneral: Domain Generalization for Left Atrial Segmentation of Multi-Center LGE MRIs}
\titlerunning{AtrialGeneral: Left Atrial Segmentation of Multi-Center LGE MRIs}

\author{Lei Li\inst{1, 2, 3} \and
Veronika A. Zimmer \inst{3} \and
Julia A. Schnabel \inst{3} \and
Xiahai Zhuang\inst{1} ${^{(\textrm{\Letter})}}$
} 
\authorrunning{L. Li et al.}


\institute{School of Data Science, Fudan University, Shanghai, China \\
\email{zxh@fudan.edu.cn} \and 
School of Biomedical Engineering, Shanghai Jiao Tong University, Shanghai, China \and
School of Biomedical Engineering and Imaging Sciences, King’s College London, London, UK}

\maketitle 
\begin{abstract}
Left atrial (LA) segmentation from late gadolinium enhanced magnetic resonance imaging (LGE MRI) is a crucial step needed for planning the treatment of atrial fibrillation.
However, automatic LA segmentation from LGE MRI is still challenging, due to the poor image quality, high variability in LA shapes, and unclear LA boundary.
Though deep learning-based methods can provide promising LA segmentation results, they often generalize poorly to unseen domains, such as data from different scanners and/or sites. 
In this work, we collect 140 LGE MRIs from different centers with different levels of image quality.
To evaluate the domain generalization ability of models on the LA segmentation task, we employ four commonly used semantic segmentation networks for the LA segmentation from multi-center LGE MRIs.
Besides, we investigate three domain generalization strategies, i.e., histogram matching, mutual information based disentangled representation, and random style transfer, where a simple histogram matching is proved to be most effective.

\end{abstract}

\section{Introduction}
Radiofrequency (RF) ablation is a common technique in clinical routine for the atrial fibrillation (AF) treatment via electrical isolation.
However, the success rate of some ablation procedures is low due to the existence of incomplete ablation pattern (gaps) on the left atrium (LA).
Late gadolinium enhanced magnetic resonance imaging (LGE MRI) has been an important tool to detect gaps in ablation lesions, which are located on the LA wall and pulmonary vein (PV).
Thus, it is important to segment LA from LGE MRI for the AF treatment.
Manual delineations of the LA from LGE MRI can be subjective and labor-intensive, and automating this segmentation remains challenging.

In recent years, many algorithms have been proposed to perform automatic LA segmentation from medical images, but mostly for non-enhanced imaging modalities.
Conversely, LGE MRI has received less attention with respect to developed methods of LA segmentation to assist the ablation procedures. 
Most of the current studies on LA segmentation from LGE MRI are still based on time-consuming and error-prone manual segmentation methods \cite{journal/JACC/higuchi2018,journal/EP/njoku2018}.
This is mainly because LA segmentation methods in non-enhanced imaging modalities are difficult to directly apply to LGE MRI, due to the existence of the contrast agent and its low-contrast boundaries.
Therefore, existing conventional automated LA segmentation of LGE MRI approaches generally require hard available supporting information, such as shape priors \cite{journal/TIP/zhu2013} or additional MRI sequences \cite{journal/MedIA/li2020}.
Recently, with the development of deep learning (DL) in medical image computing, some DL-based algorithms have been proposed for automatic LA segmentation directly from LGE MRI \cite{conf/MICCAI/li2020,journal/MedIA/xiong2020}. 

\begin{figure*}[t]\center
 \includegraphics[width=0.6\textwidth]{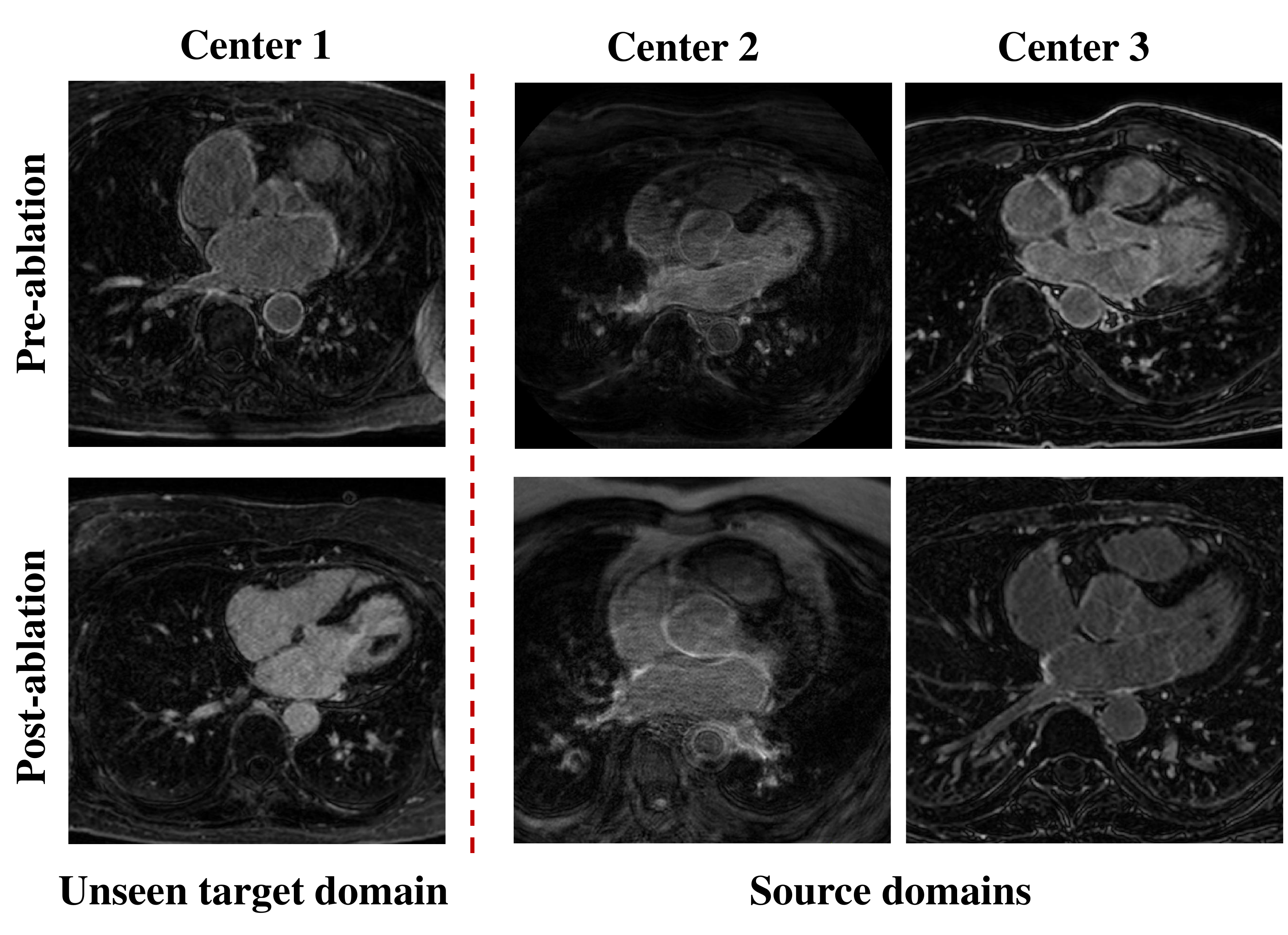}\\[-2ex]
   \caption{Multi-center pre- and post-ablation LGE MRIs. 
   The images differ in contrast, enhancement and background.}
\label{fig:AtrialJSQnet:intro:multi-center LGE MRIs}
\end{figure*}

However, the generalization ability of the DL-based models is limited, i.e., the performance of a trained model on the known domain (source domain) will be degraded drastically on an unseen domain (target domain).
This is mainly due to the existence of a \textit{domain shift} or \textit{distribution shift}, which is common among the data collected from different centers and vendors, as shown in \Leireffig{fig:AtrialJSQnet:intro:multi-center LGE MRIs}.
In the clinic, it is impractical to retrain a model each time for the data collected from new vendors or centers.
Therefore, improving the model generalization ability is important to avoid the need of retraining.
Current domain generalization (DG) methods can be categorized into three types: 
(1) domain-invariant feature learning approaches, such as disentangled representation \cite{journal/TMI/meng2020}; 
(2) model-agnostic meta-learning algorithms, which optimize on the meta-train and meta-test domain split from the available source domain \cite{conf/NIPS/dou2019}; 
(3) data augmentation strategies, which increase the diversity of available data \cite{journal/FCM/chen2020}. 

\begin{figure*}[t]\center
 \includegraphics[width=0.72\textwidth]{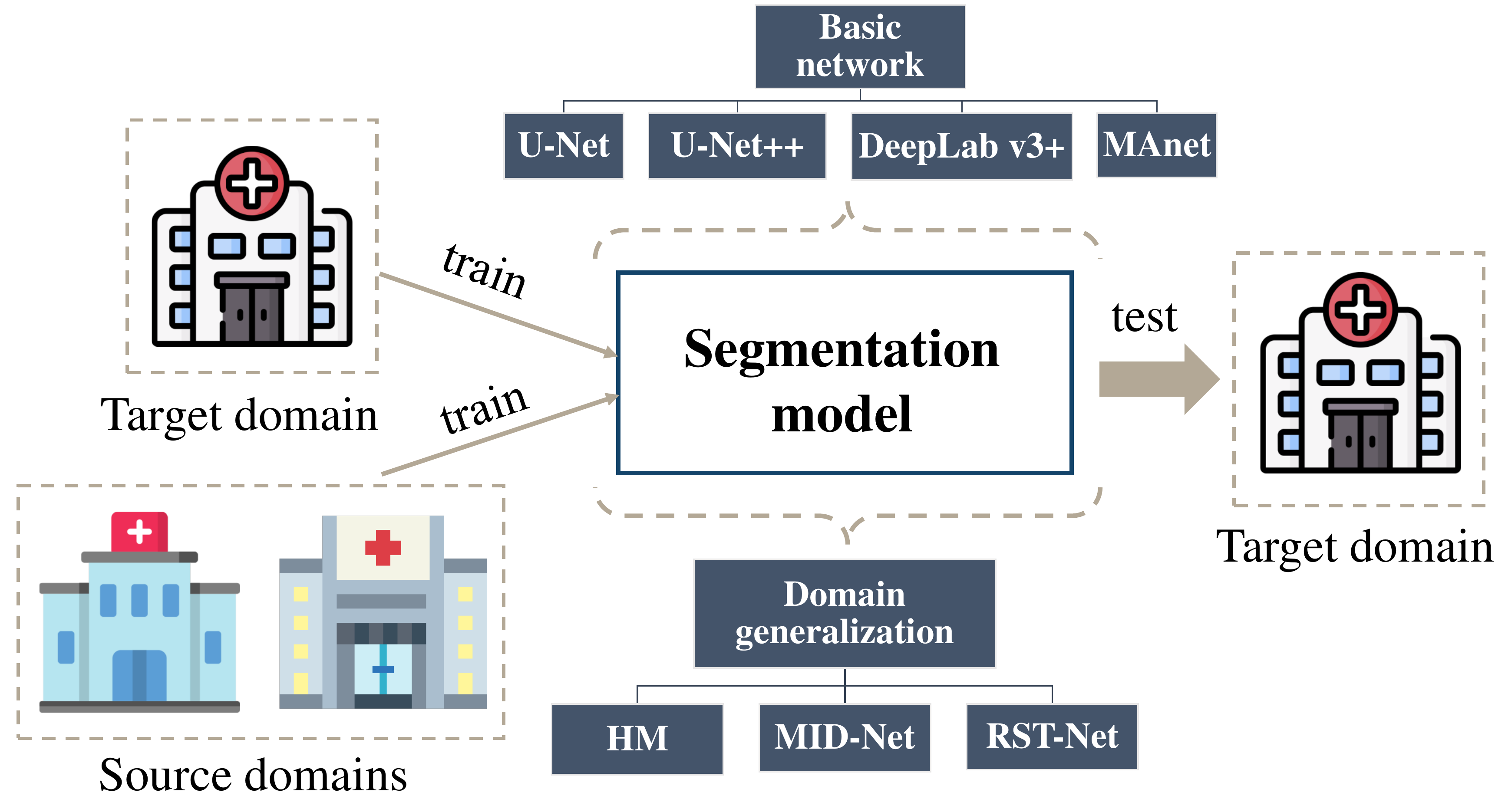}\\[-2ex]
   \caption{Illustration of the LA segmentation models for multi-center LGE MRIs.}
\label{fig:AtrialJSQnet:method}
\end{figure*}

In this work, we investigate the generalization abilities of four commonly used segmentation models, i.e., U-Net \cite{conf/MICCAI/ronneberger2015}, UNet++ \cite{conf/MICCAI/zhou2018}, DeepLab v3+ \cite{conf/ECCV/chen2018} and multi-scale attention network (MAnet) \cite{conf/CVPR/chen2016}.
As \Leireffig{fig:AtrialJSQnet:method} shows, we select two different sources of training data, i.e., target domain (TD) and source domains (SD) to evaluate the model generalization ability.
Besides, we compare three different DG schemes for LA segmentation of multi-center LGE MRIs.
The schemes include histogram matching (HM) \cite{conf/STACOM/ma2020}, mutual information based disentangled (MID) representation \cite{journal/TMI/meng2020}, and random style transfer (RST) \cite{conf/STACOM/li2020,conf/ECCV/zhou2020}.



\section{Methodology}
In this section, we describe the segmentation models we employ, formulate the DG problem (illustrated in Fig. \ref{fig:AtrialJSQnet:method}) and describe the investigated three DG strategies.

\subsection{Image Segmentation Models}
All our segmentation models are supervised approaches based on convolutional neural networks. Typically, the models are trained using a training database $\mathcal{T_\mathcal{D}}=\{(X_m, Y_m), m=1,\dots, M\}$ with images $X\in\mathcal{D}$ from a single domain $\mathcal{D}$ and corresponding labels $Y$. 
The segmentation model $f(X)$ can be defined as,
\begin{equation}
    f(X) \rightarrow Y, X \in \mathcal{D},
\end{equation}
where $X,Y \in \mathbb{R}^{1 \times H \times W}$ denote the image set and corresponding LA segmentation set.

We consider four commonly used segmentation models, all with an encoder-decoder architecture. 
The first model is a vanilla U-Net. 
U-Net++ is a modified version of the U-Net with a more complex decoder. 
DeepLab v3+ employs atrous spatial convolutions, and MAnet introduce multi-scale attention blocks.

\subsection{Domain Generalization Models}
The generalization ability of such models is limited, i.e., a model trained on a source domain $\mathcal{D}$ might perform poorly for images $X\notin\mathcal{D}$. 
DG strategies are therefore proposed to generalize models to unseen (target) domains. 
Given $N$ source domains $\mathcal{D}_s=\left\{\mathcal{D}_{1}, \mathcal{D}_{2}, \cdots, \mathcal{D}_{N}\right\}$, we aim to construct a DG model $f^{DG}(X)$,  
\begin{equation}
    f^{DG}(X) \rightarrow Y, X \in \mathcal{D}_s \cup \mathcal{D}_{t},
\end{equation}
where $\mathcal{D}_{t}$ are unknown target domains.

We investigate three DG strategies for LA segmentation from LGE MRI. 
In the first and simplest approach, HM is performed on the images from the target domain to match its intensity histogram onto that of the source domains. 
The model and training process do not change. 
The second (MID-Net \cite{journal/TMI/meng2020}) and third method (RST-Net \cite{conf/STACOM/li2020,conf/ECCV/zhou2020}) are state-of-the-art methods employing different approaches to achieve DG. 
In MID-Net, domain-invariant features are extracted by mutual information based disentanglement in the latent space, while in RST-Net available domains are augmented via pseudo-novel domains.





\section{Materials}

\subsection{Data Acquisition and Pre-processing}
LGE MRIs with various image qualities, types and imaging parameters were collected from three centers, as \Leireftb{tb:AtrialGeneral:data info} shows. 
The centers consist of Utah School of Medicine (Center 1), Beth Israel Deaconess Medical Center (Center 2), and Imaging Sciences at King’s College London (Center 3).
The dataset were selected from two public challenge, i.e., \textit{MICCAI 2018 Atrial Segmentation Challenge} \citep{link/LAseg2018} and \textit{ISBI 2012 Left Atrium Fibrosis and Scar Segmentation Challenge} \citep{link/LAScarSeg2012}.
A total of 140 images were collected and acquired either pre- or post-ablation.
The acquisition time of pre-ablation scans varied slightly among 1 to 7 days, but that of post-ablation had a range from 1 to 27 months depending on the imaging center.

The LGE MRIs from center 1, 2, 3 and 4 were reconstructed to 0.625 $\!\times\!$ 0.625 $\!\times\!$ 1.25 mm, (0.7-0.75) $\!\times\!$ (0.7-0.75) $\!\times\!$ 2 mm, 0.625 $\!\times\!$ 0.625 $\!\times\!$ 2 mm, and 0.625 $\!\times\!$ 0.625 $\!\times\!$ 1.25 mm, respectively.
All 3D images were divided into 2D slices as network inputs and then were cropped into a unified size of 192 $\!\times\!$ 192 centering at the heart region, with a intensity normalization via Z-score. 
Random rotation, random flip and Gaussian noise augmentation were applied during training.
The data distribution in the subsequent experiments is presented in \Leireftb{tb:AtrialGeneral:data distribution}.

\begin{table*} [t] \center
    \caption{
    Image acquisition parameters of the multi-center LGE MRIs.
     }
\label{tb:AtrialGeneral:data info}
{\small
\begin{tabular}{p{1.8cm}|p{3.3cm}|p{3.2cm}|p{3.2cm}} 
\hline
Parameters        & Center 1 & Center 2 & Center 3 \\
\hline
No. subject &100 & 20 & 20  \\
\hdashline
Scanner & 1.5T Siemens Avanto; 3T Siemens Vario & 1.5T Philips Achieva & 1.5T Philips Achieva  \\
\hdashline
Resolution &1.25 $\!\times\!$ 1.25 $\!\times\!$ 2.5 mm  & 1.4 $\!\times\!$ 1.4 $\!\times\!$ 1.4 mm & 1.3 $\!\times\!$ 1.3 $\!\times\!$ 4 mm  \\
\hdashline
TI, TE/TR & 270-310 ms, 2.3/5.4 ms & 280 ms, 2.1/5.3 ms  & 280 ms, 2.1/5.3 ms  \\
\hdashline
Pre-scan  & N/A  & \textless 7 days & \textless 2 days \\
\hdashline
Post-scan & 3-27 months & 30 days & 3-6 months \\
\hline
\end{tabular} }\\
\end{table*}

\subsection{Gold Standard and Evaluation}
All the LGE MRIs were manually delineated by the experts from the corresponding centers.
The manual LA segmentation were regarded as the gold standard.
For LA segmentation evaluation, Dice score, average surface distance (ASD) and Hausdorff distance (HD) were applied. 
Each image from the three centers were assigned an image quality score by averaging the scores from two experts, mainly based on the visibility of enhancements and the existence of image artefacts (please see the Supplementary Material file).

\subsection{Implementation}
The proposed framework was implemented in PyTorch, running on a computer with 2.20GHz Intel(R) Xeon(R) E5-2630 v4 CPU and a GeForce GTX 1080 Ti GPU.
We employed the released \textit{Segmentation Models} \cite{Yakubovskiy:2019} for experiments.
All the backbones of the four semantic segmentation models are the \textit{efficientnet-b6}.
We used the Adam optimizer to update the network parameters. 
The initial learning rate was set to 5e-5 and multiplied by 0.95 every 10 epochs. 

\begin{table*} [t] \center
    \caption
    {The distribution of training dataset and test dataset of LGE MRI from the three centers
    (C-$i$: center $i$).}
\label{tb:AtrialGeneral:data distribution}
{\small
\begin{tabular}{  l| l l l *{4}{@{\ \,} l }}\hline
Source of training data & Post/Pre  & \quad No. training data & \quad No. test data\\
\hline
\multirow{2}{*}{Target domain (TD)}           &Post & \quad 20 C-1 & \quad 40 C-1\\
~                                        &Pre  & \quad 20 C-1 & \quad 20 C-1\\
\hline
\multirow{2}{*}{Source domains (SD)} \quad &Post & \quad 10 C-2 + 10 C-3 & \quad 40 C-1\\
~                                        &Pre  & \quad 10 C-2 + 10 C-3 & \quad 20 C-1\\
\hline
\end{tabular} }\\
\end{table*}

\section{Experiment}
\subsection{Comparisons of Different Semantic Segmentation Networks}
\Leireftb{tb:AtrialGeneral:networks} summarizes the LA segmentation results in terms of Dice, ASD and HD based on the four semantic segmentation models.
One can see that all the segmentation models had a performance decrease when the target domain was not included in the training data.
It proves that the generalization capabilities of currently commonly used DL-based segmentation models are still very limited.
When we observe the Dice value of the LA segmentation, the obtained performances of the four models training on the TD are very close. 
However, DeepLab v3+ achieved significantly better ASD and HD than the other three models.
It may be attributed to its atrous convolution and spatial pyramid pooling module, which promote the network to learn more spatial information.
When training on the SD, the performance decrease of was DeepLab v3+ was smaller than other three models.
Therefore, in this work DeepLab v3+ is regarded as the baseline model, and we will improve its generalization ability using the proposed DG schemes.


\begin{table*} [t] \center
    \caption{
    Performance of the four segmentation models on the multi-center LGE MRI for LA segmentation.
    The training and test data distribution refer to Table~\ref{tb:AtrialGeneral:data distribution}, i.e., test data is from Center 1 dataset, while training data is from the TD (C-1 dataset) and SD (C-2, 3 dataset), respectively.
    }
\label{tb:AtrialGeneral:networks}
{\small
\begin{tabular}{  l l | l l l l l *{7}{@{\ \,} l }}
\hline
Source & \quad Matrix & \quad U-Net \cite{conf/MICCAI/ronneberger2015} & \quad U-Net++ \cite{conf/MICCAI/zhou2018} &  \quad DeepLab v3+ \citep{conf/ECCV/chen2018} & MAnet \cite{conf/CVPR/chen2016}\\
\hline
\multirow{3}{*}{TD} & Dice   &\quad $ 0.908 \pm 0.039 $  & \quad $ 0.908 \pm 0.037 $ & \quad $ 0.900 \pm 0.041 $ & $ 0.904 \pm 0.057 $\\
~                   & ASD (mm)  &\quad $ 1.35 \pm 0.722 $  & \quad $ 1.26 \pm 0.577 $ &\quad $ 1.31 \pm 0.557 $ & $ 1.37 \pm 0.849 $\\
~                   & HD (mm) &\quad $ 36.6 \pm 12.3 $  & \quad $ 28.5 \pm 14.9 $ &\quad $ 15.3 \pm 5.78 $ & $ 32.7 \pm 12.6 $\\
\hline
\multirow{3}{*}{SD}& Dice   &\quad $ 0.652 \pm 0.107 $  &\quad $ 0.633 \pm 0.104 $ &\quad $ 0.678 \pm 0.089 $ & $ 0.610 \pm 0.116 $\\
~                   & ASD (mm)  &\quad $ 6.86 \pm 1.51 $  &\quad $ 7.12 \pm 1.60 $ &\quad $ 6.19 \pm 1.65 $ & $ 8.26 \pm 1.98 $\\
~                   & HD (mm) &\quad $ 48.2 \pm 7.93 $  & \quad$ 51.1 \pm 7.62 $ &\quad $ 42.7 \pm 8.62 $ & $ 53.3 \pm 7.45 $\\
\hline
\end{tabular} }\\
\end{table*}

\subsection{Comparisons of Post- and Pre-ablation LGE MRI}
As \Leireffig{fig:AtrialJSQnet:intro:multi-center LGE MRIs} shows, the pre- and post-ablation LGE MRI can have high variability of tissue appearance.
There are already several studies that have shown the performance of LA scar segmentation and quantification varied among pre- and post-ablation LGE MRI \cite{journal/jcmr/Karim2013}. 
This is mainly because that when comparing to post-ablation images, the scars on pre-ablation LGE MRIs are hard to distinguish even for experts.
In contrast, as far as we know, there are to this date no studies comparing the LA segmentation performance for pre- and post-ablation LGE MRI.
Here, we compared and analyzed the LA segmentation performance on pre- and post-ablation LGE MRI on the four basic segmentation models.

\begin{figure*}[t]\center
	\subfigure[] {\includegraphics[width=0.47\textwidth]{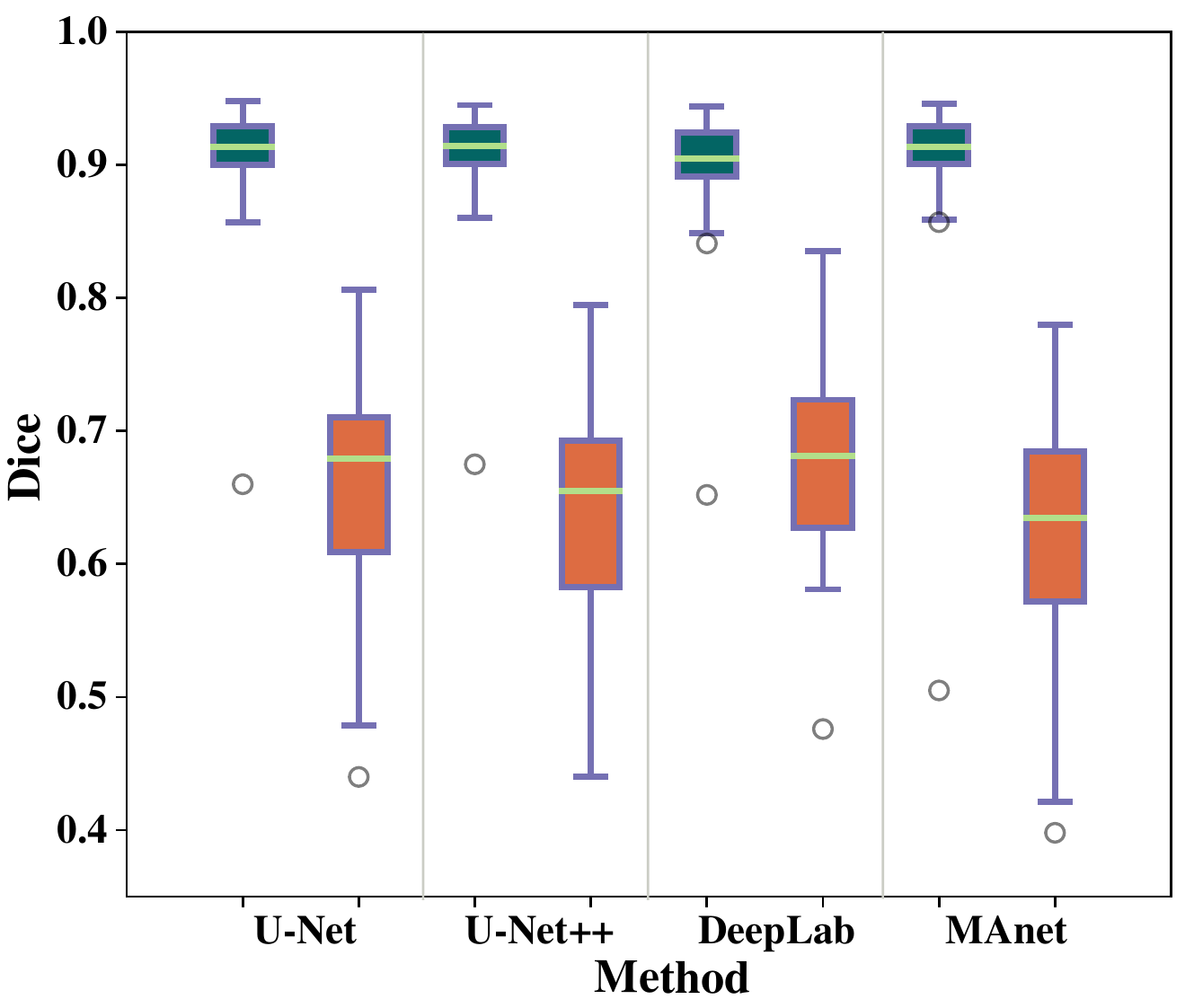}}
	\subfigure[] {\includegraphics[width=0.47\textwidth]{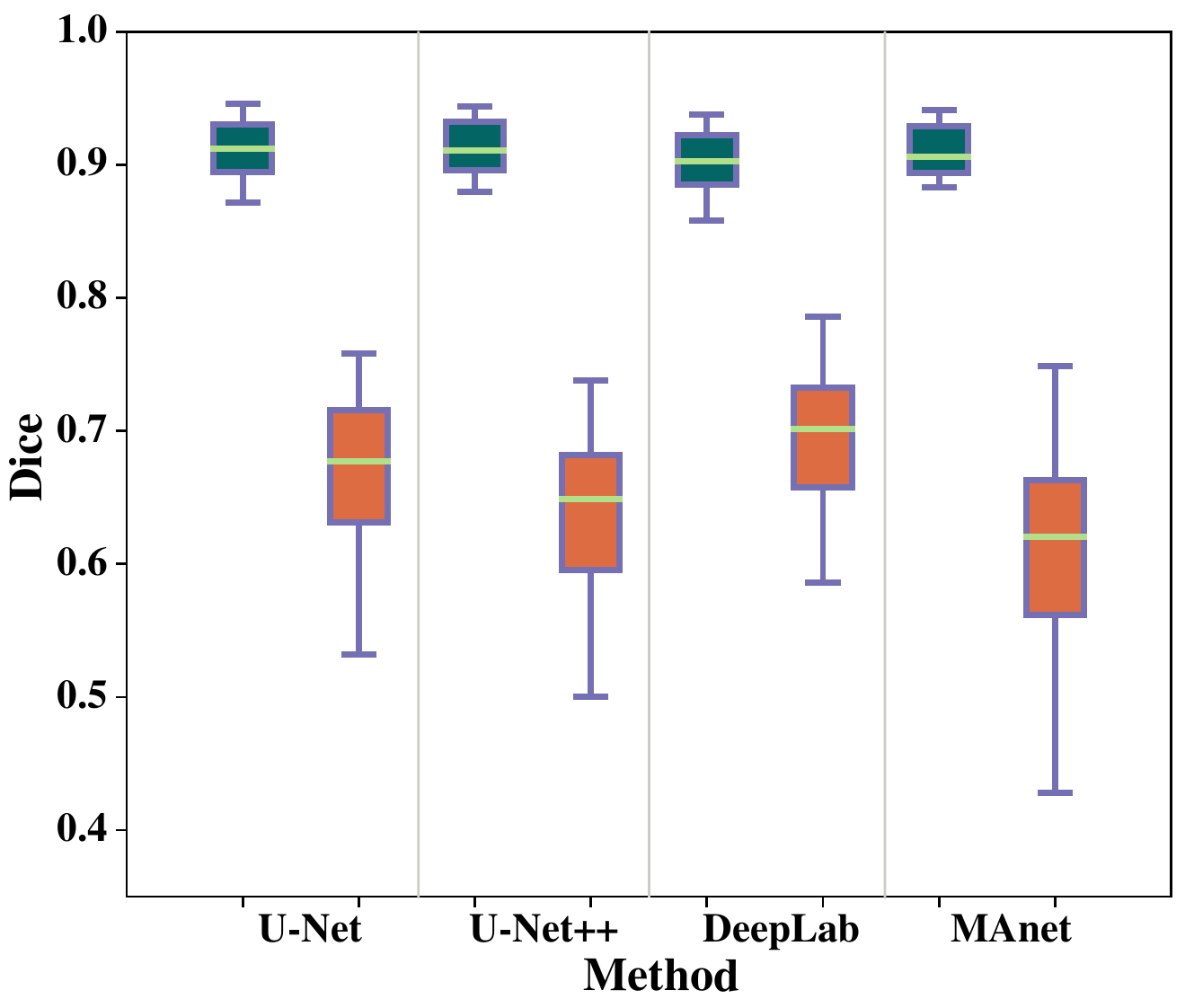}}
	\subfigure[] {\includegraphics[width=0.47\textwidth]{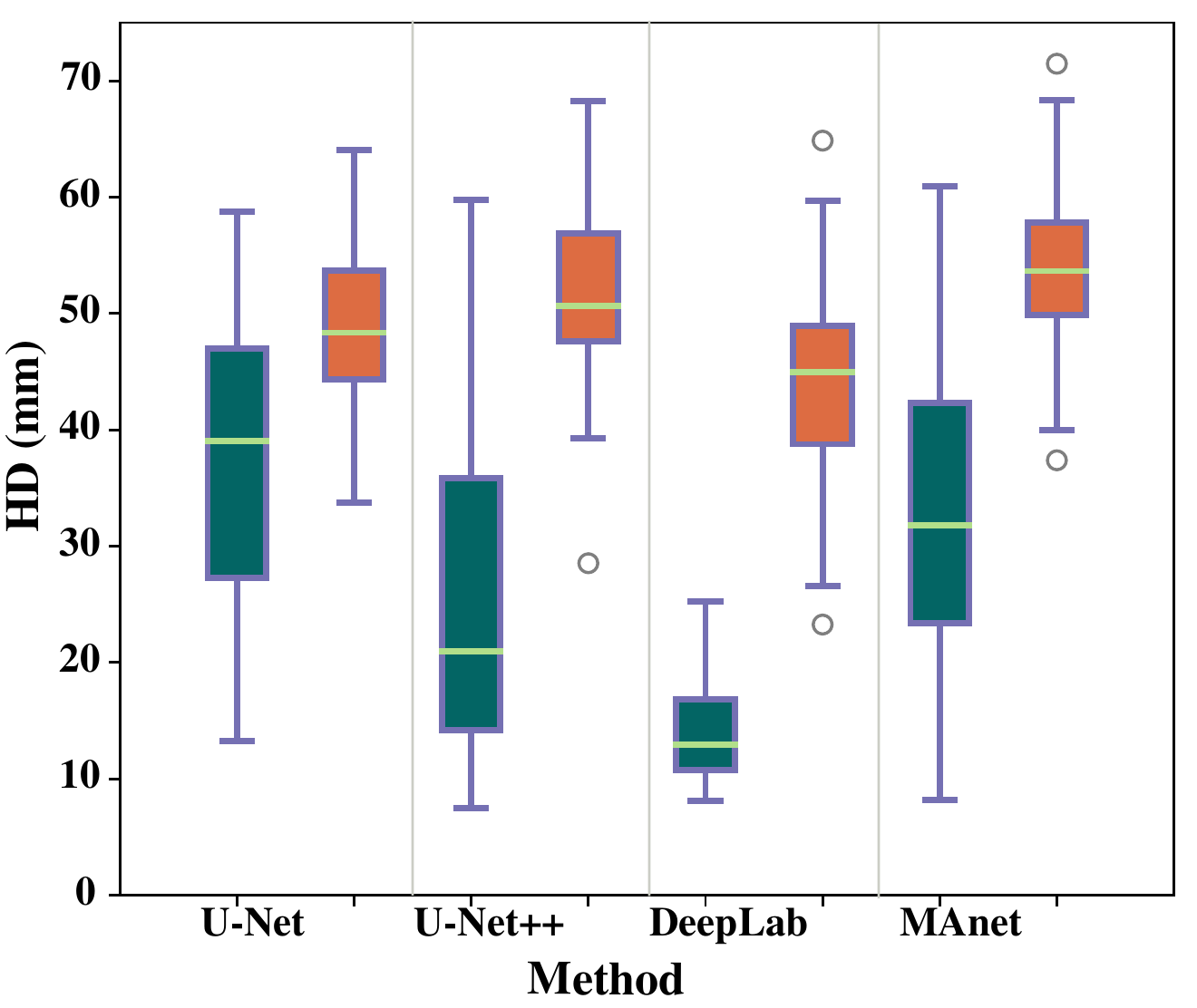}}
	\subfigure[] {\includegraphics[width=0.47\textwidth]{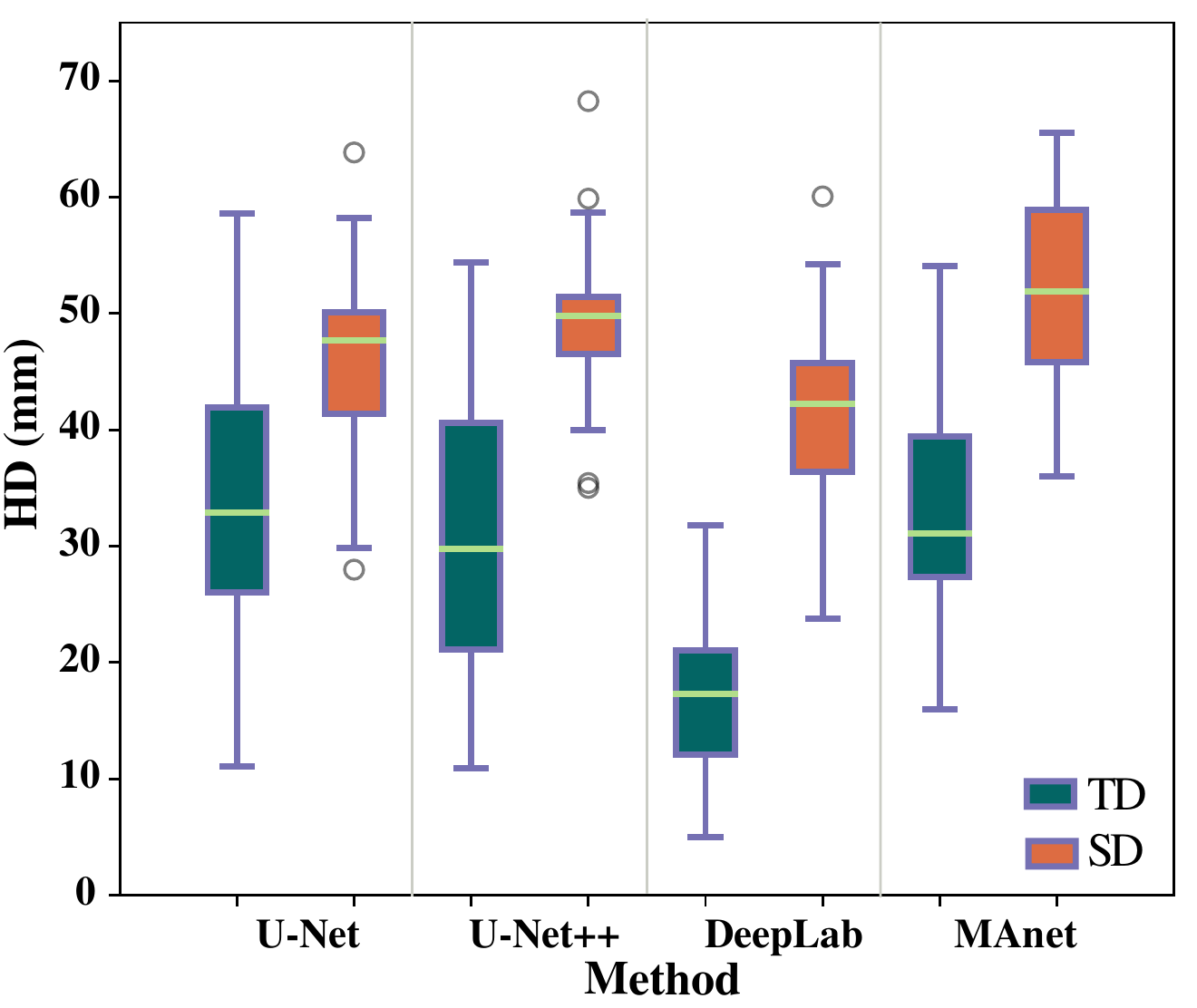}}
  \caption
    {LA segmentation results of four segmentation models with different source of training datasets:
     (a) Dice of post-ablation cases;
     (b) Dice of pre-ablation cases;
     (c) HD of post-ablation cases;
     (d) HD of pre-ablation cases.}
\label{fig:AtrialGeneral:exp:net:boxplots}
\end{figure*}

\Leireffig{fig:AtrialGeneral:exp:net:boxplots} presents the Dice and HD value obtained by the four models on the pre- and post-ablation images, separately.
One can see that, the four models all suffered from an accuracy deterioration caused by the domain shift on both pre- and post-ablation LGE MRIs, which is consistent with the results in \Leireftb{tb:AtrialGeneral:networks}.
Besides, the Dice obtained by the four models is similar on both pre- and post-ablation LGE MRIs, but DeepLab v3+ performed better in terms of HD, especially on pre-ablation data. 

In summary, there is no evident performance difference between pre- and post-ablation data for the four models.
However, the standard deviations of the Dice and HD values of the LA segmentation on the pre-ablation data are generally lower than those of post-ablation images.
It may indicate that the segmentation model is more robust for the pre-ablation data of the multi-center LGE MRIs.

\subsection{Comparisons of Different Generalization Models} \label{exp:DG}
\Leireftb{tb:AtrialGeneral:exp:DG} summarized three DG schemes to compare with the baseline DeepLab v3+ model training on multi-source domains.
One can see that three tested generalization strategies worked when comparing with baseline results.
Among the three methods, the conventional histogram matching algorithm performed best.
The MID-Net and RST-Net obtained similar results in terms of Dice, but the ASD and HD of MID-Net were worse.

\begin{figure*}[t]\center
 \includegraphics[width=0.98\textwidth]{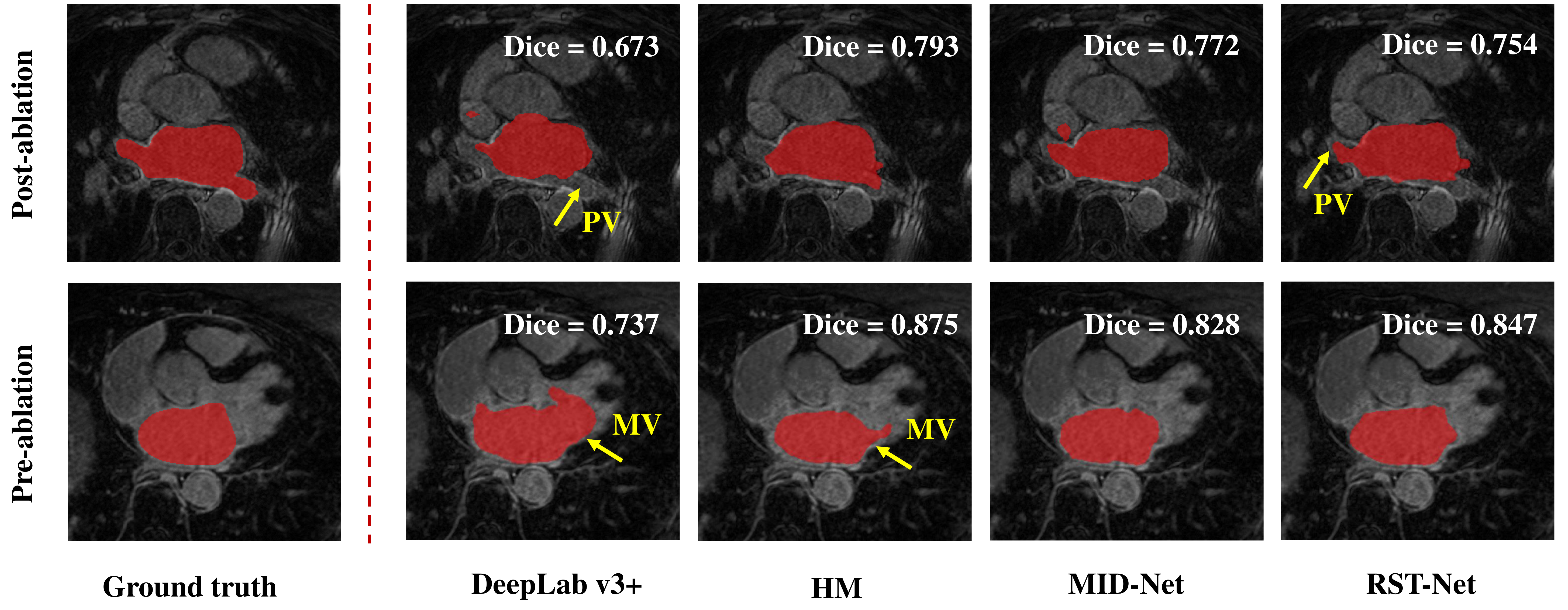}\\[-2ex]
   \caption
   {2D visualization of the LA segmentation based on the three generalization models on the multi-center LGE MRIs.
   Here, the yellow arrows indicate the wrong LA segmentation regions, i,e. PV and MV.}
\label{fig:AtrialJSQnet:exp:2d visual (models)}
\end{figure*}

\Leireffig{fig:AtrialJSQnet:exp:2d visual (models)} presents the 2D visualization results of the four methods on post-/pre-ablation LGE MRI.
In the post-ablation case, three DG schemes could identify some missing PV regions by the DeepLab v3+.
Similarly, in the pre-ablation subject, MID-Net and RST-Net both mitigated the segmentation errors in the mitral valve (MV) area.
It proved that for the both post- and pre-ablation cases, the employed DG methods worked.

\begin{table*} [t] \center
    \caption
    {Performance of different generalization models training on multi-source domains for LA segmentation.}
\label{tb:AtrialGeneral:exp:DG}
{\small
\begin{tabular}{  l | l l l *{4}{@{\ \,} l }}
\hline
Method &\quad Dice &\quad ASD (mm) &\quad HD (mm)\\
\hline
DeepLab v3+ (baseline) &\quad $ 0.678 \pm 0.089 $ &\quad $ 6.19 \pm 1.65 $ &\quad $ 42.7 \pm 8.62 $ \\
\hdashline
HM \cite{conf/STACOM/ma2020} &\quad $ 0.772 \pm 0.089 $ &\quad $ 3.76 \pm 1.04 $ &\quad $ 26.5 \pm 5.30 $ \\
MID-Net \cite{journal/TMI/meng2020} &\quad $ 0.741 \pm 0.064 $ &\quad $ 4.83 \pm 1.12 $ &\quad $ 42.6 \pm 9.66 $ \\
RST-Net \cite{conf/STACOM/li2020,conf/ECCV/zhou2020} &\quad $ 0.756 \pm 0.090 $ &\quad $ 4.20 \pm 1.15 $ &\quad $ 30.3 \pm 6.90 $ \\
\hline
\end{tabular} }\\
\end{table*}

\section{Conclusion}
In this work, we first investigated the generalization abilities of different semantic segmentation models for LA segmentation from multi-center LGE MRIs.
The results showed that all the performance of the commonly used segmentation models degraded dramatically on the unknown domain.
It emphasized the importance of promoting deep models with efficient inherent generalization abilities for LGE MRI data processing from different centers.
We then introduced three DG strategies, which were all able to alleviate the performance decrease.
Our study found that, quite surprisingly, the simple histogram matching strategy is the most effective method for DG on the LA segmentation of multi-center LGE MRI data.
It may indicate that there is still large scope for further algorithmic developments in DG.
In future, we will find the inherent differences of multi-center LGE MRIs, and develop a targeted and effective DG strategy to solve this problem. 
Moreover, we will further study the domain shift between post- and pre-ablation LGE MRI from the same center, and the label variations of LGE MRIs from different centers.


\subsubsection{Acknowledgement.}
This work was funded by the National Natural Science Foundation of China (grant no. 61971142, 62111530195 and 62011540404) and the development fund for Shanghai talents (no. 2020015).
L. Li was partially supported by the CSC Scholarship. 
JA Schnabel and VA Zimmer would like to acknowledge funding from a Wellcome Trust IEH Award (WT 102431), an EPSRC programme grant (EP/P001009/1), and the Wellcome/EPSRC Center for Medical Engineering (WT 203148/Z/16/Z).

\bibliographystyle{splncs04}
\bibliography{A_refs}

\end{document}